
\magnification=\magstep1
\font\twelvebf=cmbx10 scaled \magstep1
\hsize 15.2true cm
\vsize 22.0true cm
\voffset 1.5cm
\nopagenumbers
\headline={\ifnum \pageno=1 \hfil \else\hss\tenrm\folio\hss\fi}
\pageno=1
 
\hfill IUHET-332

\hfill  May 21, 1996
\medskip
\bigskip
\centerline{\twelvebf Spin dependence of the masses
of heavy baryons} 
 
\medskip
\bigskip
 
\centerline{D. B. Lichtenberg} 
 
\centerline{Physics Department, Indiana University, 
Bloomington, IN 47405, USA}
 
 \vskip 1 cm

It is argued from the systematics of spin-depen\-dent
forces between quarks that two proposed baryon states, 
named $\Sigma_c(2380)$ and $\Sigma_b(5760)$, do not exist. 


\vskip 1cm
\medskip \bigskip \bigskip 
 
Recently, Falk [1] has proposed that there exist
two undiscovered heavy bary\-ons: a $\Sigma_c(2380)$ that
decays radiatively to the $\Lambda_c$, and a
$\Sigma_b(5760)$ that decays radiatively to
the $\Lambda_b$. In this note I use
the systematics of the spin-dependent forces between
quarks to argue against the existence of these
two new states. The same ideas suggest that 
two of the proposed
``equal spacing rules'' [1] among heavy baryon mass 
differences should be replaced by inequalities.

Tensor and 
spin-orbit forces do not contribute perturbatively to
the masses of ground-state baryons 
and so I confine myself to baryons without 
radial or orbital excitations. 
Then only the spin-spin interaction 
(sometimes called the color-magnetic or color-hyperfine
interaction) survives in the perturbative approximation.
More detailed discussions of these points have been
given previously [2,3].

In addition to treating baryon mass splittings,
I also consider the spin-depen\-dent splittings
of meson masses, as under certain assumptions there
are inequalities relating meson and baryon mass
differences.
I neglect mass splittings among isospin multiplets,
which means neglecting electromagnetic effects and 
the mass difference between the $d$ and $u$ quark 
Also, as is often done, I
let the symbol for a hadron denote its mass, averaged
over isospin states if more than one exists. 

In order to motivate certain inequalities among 
matrix elements of the spin-spin interaction, it is 
convenient to assume [2,3] that the form of the spin-spin
interaction $I_S$ between quarks in a baryon or a quark
and an antiquark in a meson has the form
$$I_S=- 3\sum_{i<j}\lambda_i\cdot \lambda_j\  
\sigma_i\cdot\sigma_j\ f(r_{ij})/
(8m_im_j), \eqno(1)$$
where $i$ and $j$ denote quarks (or antiquarks), 
the $\lambda_i$ and $\lambda_j$ are color Gell-Mann
matrices, the $\sigma _i$ and $\sigma_j$ are Pauli spin
matrices, and $f(r_{ij})$ is a positive definite
function of the distance between the two particles. 
The factor 3/8 is chosen for convenience. 
In the Fermi-Breit approximation to one-gluon exchange
in QCD, the spin-spin interaction is a special
case of this form, namely 
$$f(r_{ij}) = 16\pi 
\delta(r_{ij})/9 , \eqno(2)$$

It should be stressed that it
is not necessary to assume the validity of Eq.\ (1).
The inequalities among hadron masses
may alternatively be obained from the systematics
of the observed spin splittings of 
hadrons containing only light quarks. 
The essential point of this paper
is that it is very reasonable that the inequalities
should also hold for hadrons containing  heavy quarks.

The expectation values for the color and spin 
operators in $I_S$
can be taken explicitly [3]. The expectation
value of the spatial operator can be given in terms
of quantities $R_{ij}$ for mesons and $R_{ijk}$,
$R_{ikj}$, and $R_{kji}$ for baryons [3].
For mesons,
$$R_{ij}= 2\langle ij | f(r_{ij}) |ij \rangle /
(m_im_j), \eqno(3)$$
where $|ij\rangle $ is the unperturbed meson spatial
wave function, For baryons
$$R_{ijk}=\langle ijk| f(r_{ij})|ijk \rangle /
(m_im_j), $$
$$R_{ikj}=\langle ijk| f(r_{ik})|ijk \rangle /
(m_im_k), \eqno(4)$$
$$R_{jki}=\langle ijk| f(r_{jk})|ijk \rangle /
(m_jm_k), $$
where $|ijk\rangle$ is the unperturbed baryon spatial
wave function. The ordering of the quarks in $|ijk
\rangle$ is important here [4]. If all three quarks
are different, the two lightest are the first two;
if two are identical, these are the first two.
Although the operator $f(r_{ij})$ for baryons 
is a two-quark
operator, the expectation values $R_{ijk}$ etc.
depend also on the third or ``spectator'' quark
through the 3-quark wave function.
The $R_{ijk}$ are symmetric under the 
interchange of their first two indices:
$$R_{ijk}= R_{jik},\eqno(5)$$ 
but in general (all quarks different)
$$R_{ijk}\not=  R_{ikj}\not= R_{jki}\eqno(6)$$ 
However, because of the neglect of the mass 
difference between the $u$ and $d$ quark, if $i$ is
a $u$ quark and $j$ is a $d$ quark, then
$R_{ukd}=R_{dku}$. From here on, I denote both
$u$ and $d$ quarks by the symbol $q$.

Let $M^*_{12}$ and $M_{12}$ denote ground 
state vector and pseudoscalar mesons containing quark
1 and antiquark 2. For baryons, if all three quarks
have different flavors, let 
$B^*_{123}$ denote the baryon
of spin 3/2 containing containing quarks 1, 2, and 3, and
$B_{123}$, and $B'_{123}$ denote two different spin 
1/2 baryons. The baryons $B$ and $B'$ are distinguished
by the spin of the two lightest quarks 1 and 2;
in $B$ these quarks have spin 0 and in $B'$ they have 
spin 1. If two quarks in the baryon have the same 
flavor, they
are 1 and 2, and the state $B$ is absent. If all three
quarks have the same flavor, both $B$ and $B'$ are absent.

The $R_{ij}$ and $R_{ijk}$ contribute to the masses of 
these ground-state mesons and baryons as follows [3]:   
$$M^*_{12} = E_{12} + R_{12}, \quad
M_{12} = E_{12} - 3 R_{12}, \eqno(7)$$
$$B^*_{123} = E_{123} +  R_{123} +R_{132}
+ R_{231}, $$
$$B'_{123} = E_{123} +  R_{123} -2R_{132}
-2 R_{231}, \eqno(8)$$
$$B_{123} = E_{123} - 3  R_{123}, $$
where $E_{12}$ and $E_{123}$ are the meson and baryon
masses in the absence of the spin-dependent force.

The form of Eqs.\ (3), (4)  suggest the following
inequalities [2] among the $R_{ij}$ and $R_{ijk}$:
$$R_{ij} > R_{il}, \ \ R_{ijk} > R_{iln} \ \ 
{\rm if} \ \ m_j < m_l, \eqno(9)$$
because the quark masses appear in the denominator in
Eqs.\ (3), (4). Furthermore, because the expression
for $R_{ij} $ in Eq.\ (3) contains a factor 2 compared
to the expression for $R_{ijk}$ in Eq.\ (4),
it is plausible that 
$$R_{ij} >2 R_{ijk}. \eqno(10)$$
Likewise, it is plausible that the inequality
$$R_{ijk} < R_{ijl} \ \ 
{\rm if} \ \ m_k < m_l, \eqno(11)$$
holds [2].
The inequalities (10) and (11) should be valid for
any function $f(r_{ij})$ of sufficiently  short range,
as the following argument shows:
Because the short-range
part of the quark-antiquark potential (arising
from one-gluon exchange) is twice as large for
mesons as for baryons, the meson wave function
is not as spread out in space as the baryon wave
function. Because both wave functions
are normalized to unity,
the meson wave function must be larger than the
baryon wave function at small spatial separations
where $f(r_{ij})$ is large, so that
$$\langle ij| f_(r_{ij})|ij \rangle> 
\langle ijk| f_(r_{ij})|ijk\rangle . \eqno(12)$$
Using this
result in Eqs.\ (3) and (4), I obtain the inequality (10).
The inequality (11) follows
from the fact that for potentials like the
quark-quark potential, the radial extent of a two-particle
wave function decreases as the reduced mass 
increases. This is principally a kinematic effect.
If the mass of the spectator quark $k$ is increased, 
the quarks $i$ and $j$ are pulled closer to $k$
and consequently to each other. This reduces the
radial extent of the wave function and 
therefore increases
$\langle ijk| f_(r_{ij})|ijk \rangle$, 
from which (11) follows. However,
in the limit of heavy quark effective theory, 
i.e., $k$ and $l$ are considered to be infinitely heavy, 
the inequality (11) becomes an equality [1] because the
reduced mass of either quark $i$ or $j$ is just its
actual mass. 

From Eqs.\ (3) and (4), the $R$'s can be obtained
in terms of the observed hadron masses:
$$R_{12}= (M^*_{12} - M_{12})/4, \eqno(13)$$
$$R_{123}= (2B^*_{123} + B'_{123}- 3B_{123})/12, 
\eqno(14)$$
$$(R_{132} + R_{231})/2 = (B^*_{123} - B'_{123})/6.
\eqno(15)$$
If quarks 1 and 2 are either identical or are 
$u$ and $d$ quarks respectively, then $R_{132}= R_{231}$. 
Now Eqs.\ (13)--(15) may be taken as the definition
of the $R$'s, and the inequalities (9)--(11) may
be postulated to hold independently of
the validity of the interaction (1). The
inequalities may then be tested with the data.

In Ref.\ [3], the observed meson and baryon masses
were used to obtain values of some of the $R_{ij}$ and
$R_{ijk}$. It is useful to repeat this procedure 
using Eqs.\ (9)--(11) with the
new data that are available [5]--[8]. 
It should be noted that some of the new baryon
data are preliminary.

For mesons, I use the data from the Particle Data
group [5]. The results for mesons are (in MeV)
$$R_{qq}= 158, \quad R_{qs} = 99.5, $$
$$R_{qc}= 35.5, \quad R_{sc} = 35.4, \quad
R_{cc}= 29.2, \eqno(16)$$
$$ R_{qb}=11.5, $$
where the experimental errors are less than 1 MeV. 
Missing from Eqs.\ (16) is $R_{ss}$ because
neither the $\eta$ nor the $\eta'$ is a pure $\bar s s$
state. Also missing are $R_{cb}$ and $R_{bb}$ because
of the absence of data. These results have changed
little from those given in Ref.\ [3]. These values
of $R_{ij}$ satisfy all the meson inequalities, except
that, within the errors, $R_{qc}= R_{sc}$. This fact
may indicate that the shrinking of the wave function
when  $q$ is replaced by $s$ compensates for the
replacement of $m_q$ by $m_s$ in the denonominator
of $R_{ij}$. 

For baryons, 
I use the same data that Falk [1] used in his Table II,
with two exceptions.
His paper should be consulted for the experimental
references. The first exception is that
I assign the $\Xi_c'$ the mass
$2573\pm 4$ MeV [8], as the error is smaller than
the error in Falk's reference. The second exception is
that in addition to using the DELPHI data given by 
Feindt [7], as Falk does, I also compare with the
earlier DELPHI data quoted by Jarry [6].

First I use
the conventional mass assignments for $\Sigma_c$,
The results for baryons are (in MeV)
$$R_{qqq}= 48.8, \quad R_{qqs}= 51.2, \quad
R_{qsq}= 32.0, \quad R_{qss}= 35.8, $$
$$R_{qqc}= 54.8\pm 1.2, \quad R_{qcq}= 12.8\pm 1.2, \quad
R_{qqb}= 52.6\pm 2, \quad R_{qbq}= 9.3\pm 1.4, \eqno(17)$$
$$R_{qsc}= 38.1\pm 1, \quad (R_{qcs} +R_{scq})/2
= 11.8\pm 1, $$
where errors less than 1 MeV are omitted. In some
instances, I have added statistical and systematic
errors in quadrature. Other $R_{ijk}$ are missing
either because of absence of data or because
Eqs.\ (14), (15) are not sufficient to compute them.

Although in a few cases, the central values
of the $R_{ij}$, $R_{ijk}$, and
$R_{ikj}$ do not satisfy the 
inequalities (9)--(11), these quantities do 
satisfy the inequalities within the errors
except that $R_{qbq}$ is too large by about 3 
standard deviations. On the other hand, if instead
of using the DEPHI data [7] quoted by Falk, I use
the earlier DELPHI data quoted by Jarry [6], $R_{qbq}=
4.1 \pm 0.4$ MeV, a value which satisfies  the
inequalities, but $R_{qqb}=51.9\pm 2$, a value which
is in marginally greater disagreement with
the inequalities. The DELPHI data are still preliminary.
In the limit of heavy quark effective theory, 
$R_{qqc}= R_{qqb}$. It can be seen from the values
of these quantities given in Eq.\ (17) that, within
experimental error, the heavy quark limit has been
reached.

I expect that more precise experiments will confirm
all the inequalities of this paper. 
If not, it would mean that
the systematics of the spin splittings which hold
for hadrons containing only light quarks do not carry 
over to hadrons containing a 
heavy quark. It is of course premature to speculate on the
possible reason for such a hypothetical departure
of baryon masses from the regularities noted here. 

On the other hand, with Falk's new assignments [1],
I obtain  (in MeV)
$$R_{qqc}= 35.9\pm 1, \quad R_{qcq}= 12.2\pm 1, \quad
R_{qqb}= 40.2\pm 2, \quad R_{qbq}= 6.0\pm 1.4, \eqno(18)$$
The values of $R_{qqc}$ and $R_{qqb}$ grossly
violate the inequality (11).  a fact that leads me
to the conclusion that the postulated states
$\Sigma_c(2380)$ and $\Sigma_b(5760)$ do not exist.

I now turn to the mass equalities given in Falk's
paper [1]. Two of these equalities are
$$\Sigma_c^* - \Sigma_c = \Omega_c^* - \Omega_c, 
\eqno(19)$$
$$(2\Sigma_c^* + \Sigma_c)/3 -\Lambda_c =
(2\Xi_c^* + \Xi_c')/3 -\Xi_c. \eqno(20)$$ 
It can be seen from the inequalities satisfied by
the $R_{ijk}$ etc. that Eqs.\ (19) and (20) get
replaced by
$$\Sigma_c^* - \Sigma_c > \Omega_c^* - \Omega_c, 
\eqno(21)$$
$$(2\Sigma_c^* + \Sigma_c)/3 -\Lambda_c >
(2\Xi_c^* + \Xi_c')/3 -\Xi_c. \eqno(22)$$ 
The inequality (21) cannot be tested at present
because the $\Omega_c^*$ has not been observed.
The inequality (22) is satisfied with the 
conventional assignments for the $\Sigma_c$
and $\Sigma_c^*$, but violated for Falk's
assignments. 


Although I have arrived at the inequalities by
considering a spin-spin interaction of the form
(1) with short-range $f(r_{ij})$, this is really
not necessary. The observed pattern of spin 
splittings in mesons and in baryons containing
only light quarks is such as to satisfy all
the inequalities of this paper. All that is 
really needed is the assumption
that the pattern persists in heavy baryons. 

In conclusion, if the heavy baryons have their conventional 
spin assignments, then inequalities in 
spin-dependent mass splittings which are satisfied
for  hadrons containing only light
quarks are also satisfied for observed baryons
containing a heavy quark. (There is a possible exception
in one case involving $b$ baryons
at the three standard deviation level). However, 
if the heavy baryons are given Falk's new assignments,
some of the inequalities of this paper are seriously
violated not only for $b$ baryons,
where the data are preliminary, but also for 
$c$ baryons, which are better measured.

Part of this work was done at the Institute for
Nuclear Theory (INT) of the University of Washington. 
The author thanks the members of the INT for 
their hospitality and support. This work was supported
in part by the Department of Energy.
References

\bigskip

[1] A. F. Falk, talk at the Madison, Wisconsin, 
Phenomenology
Symposium (April, 1996) and ``A new interpretation of the 
observed heavy baryons,'' Johns Hopkins University report 
(1996, unpublished); hep-ph/9603389.

[2] D. B. Lichtenberg, Phys.\ Rev. D 35, 2183 (1987).

[3] M. Anselmino, D. B. Lichtenberg, and E. Predazzi,
Z. Phys.\ C 48, 605 (1990).

[4] J. Franklin, D. B. Lichtenberg, W. Namgung, and
D. Carydas, Phys.\ Rev.\ D24, 2910 (1981).

[5] Particle Data Group: L. Montanet et al., Phys.\ Rev.\ D 
50, 1173 (1994).


[6] P. Jarry in Physics in Collision 15, edited by
M. Rozanska and K. Rybicki, World Scientific, Singapore
(1996), p. 431.

[7] M. Feindt, talk given at the 6th International
Conference on Hadron Spectroscopy, Hadron '96, Manchester,
(July 9--14, 1995).

[8] M. S. Alam, talk given at Baryons '95, Santa Fe, NM
(October, 1995), proc. to be published.

\bye